\documentclass[prl, twocolumn, showpacs, floatfix, letterpaper, groupedaddress]{revtex4}
\usepackage[dvips]{graphicx}
\usepackage{amsfonts}
\begin{document}

\title[Thin vibrated granular layers]{The dynamics of thin vibrated granular layers }

\author{P Melby\dag, F Vega Reyes\dag, A Prevost\S, R Robertson\dag, P Kumar\dag, D A Egolf\dag and 
J S Urbach\dag\footnote[4]{To
whom correspondence should be addressed.}}

\address{\dag\ Department of Physics, Georgetown University, Washington, DC 20057, USA}

\address{\S\ Laboratoire des Fluides
Organis\'es, CNRS-UMR 7125, Coll\`ege de France, 11 place Marcelin
Berthelot 75231 Paris cedex 05, FRANCE}

\begin{abstract}
We describe a series of  experiments
and computer simulations on vibrated granular media in a geometry
chosen to eliminate gravitationally induced settling.  The system consists of a
collection of identical spherical particles on a horizontal plate vibrating vertically, with or without a confining lid.   Previously reported results are reviewed, including the observation of homogeneous, disordered liquid-like states, an instability to a `collapse' of motionless spheres on a perfect hexagonal lattice, and a fluctuating, hexagonally ordered state.  In the presence of a confining lid we see a variety of solid phases at high densities and relatively high vibration amplitudes, several of which are reported for the first time in this article.  The phase behavior of the system is closely related to that observed in confined hard-sphere colloidal suspensions in equilibrium, but with modifications due to the effects of the forcing and dissipation.   We also review measurements of velocity distributions, which range from Maxwellian to strongly non-Maxwellian depending on the experimental parameter values.  We describe measurements of spatial velocity correlations that show a clear dependence on the mechanism of energy injection.  We also report new measurements of the velocity autocorrelation function in the granular layer and show that increased inelasticity leads to enhanced particle self-diffusion.
\end{abstract}


% Uncomment for Submitted to journal title message
%\submitted

% Comment out if separate title page not required
\maketitle

\section{Introduction}
Despite its apparent simplicity, a granular medium consisting of
a collection of dry, cohesionless, identical spherical particles can
display a wide range of complex behaviors for which no reliable
mathematical models exist. The empirical relations used by engineers to describe
granular flows cannot
be reliably extended to deal with new processes or new environments.  Attempts to improve models of granular media have been thwarted by a range of
challenging problems. Collisions between granular particles may dissipate
a considerable fraction of the available kinetic energy, so the
results of equilibrium kinetic theory are not directly
applicable. Furthermore, dissipation often leads to the
appearance of large gradients and shocks, which makes it difficult to
employ continuum equations\cite{glasser01,tan98}.   
Another set of complications arises from the fact that
macroscopic particles settle rapidly under the Earth's gravity, so
that granular gases, in which particles interact through brief
collisions rather than enduring contacts, are difficult to produce
experimentally. When, as is most often the case, 
collisional pressures aren't sufficient to
support the weight of the granular media, a complex network of
particle contacts results.  
This leads to additional complications because of
the complexity of the
frictional inter-particle contacts. Moreover, the inability of
particles to surmount the gravitational barriers preventing one
particle from moving over another results in a kind of frozen-in
metastable state.  
This state has much in common with other glassy
states of matter, including the fact that it is not very well
understood.  

\begin{figure}[h!]
\centerline{
\includegraphics[width=7cm]{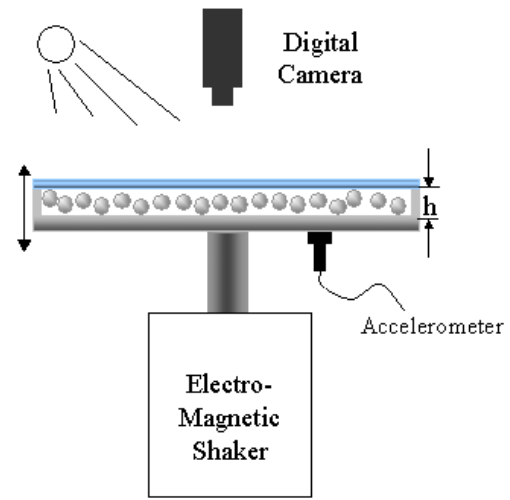}}
\caption{\small{Sketch of the experimental apparatus used in these
    studies.  In some experiments, a rough bottom plate is used in place of the
    smooth one pictured here and in others, the confining lid is absent.  In section \ref{confinement}, the effect of varying the gap height, $H$, is described. }}
\label{apparatus}
\end{figure}

In this article we describe a series  of experiments
and computer simulations on vibrated granular media in a geometry
chosen to eliminate gravitationally induced settling.  The system consists of a
collection of identical spherical particles on a horizontal plate vibrating vertically, with or without a confining lid (Fig.\ref{apparatus}). This system is
well suited to investigate many of the complex non-equilibrium effects
observed in excited granular media and is simple enough that it can be accurately modelled in molecular dynamics simulations.  By controlling the amplitude and frequency of the plate vibration, the density of spheres on the plate, the surface properties of the bottom plate, the size of the gap between the bottom plate and the confining lid, and the inelasticity of the spheres (by using different metals), we can observe a wide range of behaviors.  At low densities and moderate vibration amplitudes, a horizontally homogeneous, disordered liquid-like state is observed.  As the shaking amplitude is decreased, the granular liquid becomes unstable to a `collapse' of motionless spheres into a perfect hexagonal lattice. As described in Section \ref{phases}, this transition is probably due to a bi-stability in the interaction between the balls and the vibrating plate at low vibration amplitudes.  At higher densities we see numerous different solid phases, several of which are reported for the first time in this article.  The phase behavior of the system is closely related to that observed in confined hard-sphere colloidal suspensions in equilibrium.  The equilibrium transitions are driven by entropy maximization, and the close similarity strongly suggests a common mechanism.  However we also observe nonequilibrium effects, including non-equipartition of kinetic energy between coexisting phases\cite{prevost04}, that directly affect the phase diagram.

We have observed that the liquid-like states can have strongly non-Maxwellian velocity distributions, while in other conditions the distributions are indistinguishable from a Maxwell distribution\cite{olafsen98,olafsen99,urbach01,prevost02}.  The granular gas displays strong spatial velocity correlations, but those correlations change dramatically when the smooth horizontal plate is replaced with a rough one\cite{prevost02}.  Taken together these results demonstrate that the statistical properties of the granular gas are dependent on the external forcing that is necessary to maintain a steady state in the presence of dissipative interparticle collisions.  This effect is probably exaggerated in the very thin granular layers that we have studied because every particle frequently interacts with the confining boundaries.  Other experiments and computer simulations on quasi two-dimensional granular gasses have shown strongly non-Maxwellian velocity distributions and correlations that can be understood at least in part through consideration of the interactions with the boundaries\cite{olafsen98,olafsen99,losert99,rouyer00,blair01,aranson01,baldassarri01,vanzon04,barrat01,barrat02}.  This is consistent with  theoretical work  on the
non-equilibrium steady state obtained when energy is supplied to the granular gas by a
spatially homogeneous random external forcing\cite{vannoije98,vannoije99,puglisi99,pagonabarraga01}. Non-Maxwellian
velocity distributions and algebraically decaying velocity
correlations arise as a direct consequence of the energy
injection.  These distributions and correlations are necessary ingredients of a kinetic theory of granular fluids, and can have significant effects on transport coefficients.  As an example, in this paper we report new measurements of the velocity autocorrelation function in the granular layer, and show that increased inelasticity leads to increased correlations, which in turn enhances particle self-diffusion.

This paper is organized as follows:  Section \ref{methods} (Methods) describes the experimental apparatus and measurement techniques as well as the model used for the molecular dynamics simulations. Section \ref{phasediagram} describes the phases and phase transitions observed.  Section \ref{kinetic} reviews the statistical characterization of granular gasses.  Finally, a few concluding remarks are offered in Section \ref{conclusion}.
\section{Methods}
\label{methods}
\subsection{Experimental Apparatus}

The experimental apparatus consists of a rigid
aluminum plate approximately 20 cm in diameter that is mounted horizontally on an electromagnetic
shaker that oscillates the plate vertically.   The plate is carefully
levelled, and the peak acceleration, monitored with a fast-response accelerometer mounted on the bottom surface of the plate is uniform across the plate
to better that 0.5\%.   For all of the experiments described in this paper the particles are smooth spheres of either stainless steel (coefficient of restitution $e=0.95$) or brass ($e=0.77$,) with diameters, $\sigma$, ranging from 1.5 mm to 3 mm, depending on the experiment.   Most  experiments and simulations have been performed with a smooth plate, but in some cases we have used  a `rough plate', formed by a hexagonally close-packed layer of 1.19 mm diameter stainless steel spheres glued to the bottom plate. The particles are surrounded by an aluminum rim, and for some experiments they are confined from above with a thick, anti-static coated Plexiglas lid. 
The particles are illuminated from above to produce a small bright spot
at the top of each of them when viewed through a video camera mounted
directly above the plate.  For all of the experiments described below, the plate vibration is sinusoidal, and typically in the range of $\nu=30-100$ Hz.  The peak plate acceleration, $A\omega ^2$, where $A$ is the amplitude and $\omega = 2 \pi \nu$, typically ranges from 1 $g$ up to 5 $g$.  It is often convenient to report results in terms of the dimensionless parameter $\Gamma=A \omega ^2 /g$, 

Several digital video cameras have been used for data acquisition, including a
high-resolution camera for studying spatial correlations (Pulnix-TM1040,
1024 x 1024 pixels$^2$, 30 frames/second), and a
high-speed camera for measuring velocity distributions (Dalsa CAD1,
128 x 128 pixels$^2$, 838 frames/second).  For some experiments, we obtained high spatial and temporal resolution by using the Pulnix-1040 in combination with a stroboscopic LED array, with the strobe synchronized so that one light pulse occurs at the end of the exposure of one frame and the next pulse occurs at the beginning of the subsequent frame \cite{prevost02, prevost04}.   The most recent data has been taken with a Cooke 1200HS camera, which combines high resolution (1280x1024 pixels$^2$) with a high frame rate (625 frames/second at maximum resolution, higher for lower resolution).
The acquired digital
images are analyzed to determine particle locations by calculating
intensity weighted centers of bright spots identified in the images.

When the frame rate of camera is faster than
the inter-particle collision rate, it is possible to measure
particle velocities by measuring the displacement of the particle from
one image to the next.  The trajectories determined from the images
are nearly straight lines between collisions.  Although the frame rate
is fast compared to the mean time between collisions, the fact that it is finite
introduces unavoidable systematic errors into the experimentally
determined velocity distributions.  Some particles will undergo a
collision between frames, and the resulting displacement will not
represent the true velocity of the particle either before or after the
collision.  Since fast particles are more likely to undergo a
collision, the high energy tails of the distribution will be
decreased.  The probability that a particle of velocity $v$ will not
undergo a collision during a time interval $\Delta t$ is proportional
to exp$(-v \Delta t / l_o)$, where $l_o$ is the mean free path.  The
effect of this on the velocity distribution function reported here is
quite small.  In addition to taking velocities out of the tails of the
distribution, collisions will incorrectly add the velocities of those
particles elsewhere in the distribution.  This effect can be minimized
by filtering out the portion of particle trajectories where collisions
occur\cite{urbach01}.

\subsection{Molecular Dynamics Simulations}
While many of the quantities of interest can be gathered directly from
the experiment, some, like the vertical positions and velocities of the particles, are
very difficult to measure.   These same parameters are easy to measure using computer simulations which 
also allow systematic
variations of material parameters and can provide useful confirmation
of the interactions underlying the observed effects. 

We have written code based on simplified particle interactions
that have been used in molecular dynamics simulations of granular
media\cite{herrmann93} and have successfully reproduced, in
considerable detail, many of our experimental results\cite{nie00, prevost02, prevost04}.  
The inter-particle interactions include an elastic repulsive 
contact force that is
proportional to the particle overlap, an inelastic normal force proportional to the relative normal velocities, and a dissipative
tangential friction force proportional to the relative
tangential velocities of overlapping particles.  For computational efficiency, the particles in the simulation
are softer than the steel or brass particles used in the experiment, but the parameters give realistic coefficients of restitution.  This model does not
include a tangential restoring force that has been found to be
necessary in some instances\cite{louge00,kondic99} but may be less important
in the un-sheared systems we have studied so far.
In the case of simulations of the smooth plate, interactions with 
the top and bottom boundaries are treated like collisions with infinitely massive planes, while for the rough plate, the bottom boundary is replaced with spheres of
infinite mass.  Either periodic boundaries or elastic or inelastic side
walls can be used.

\section{Phase Diagram}
\label{phasediagram}
\subsection{Free surface layer}
 \label{phases}

This simple experimental system displays a rich phase diagram that depends on the density of particles, the plate vibration amplitude $A$, frequency $\nu$, the gap height $H$, and the coefficient of restitution $e$, of the particles.  For modest vibration amplitudes, the effect of gravity is enough to keep the layer effectively two-dimensional, and the lid does not play a significant role.  We will discuss this regime first and then describe some new results on the effect of the confining gap at higher accelerations.

At very low densities, the interparticle collisions are not frequent enough to overcome friction with the bottom plate, and the granular layer does not exhibit any significant horizontal motion.  For a range of surface coverages above about $\rho_{2D} = 25\%$\footnote{$\rho_{2D} = N/N_{max}$ where $N$ is the number of balls on the plate and $N_{max}$ is the maximum number of balls that fit on the plate in a hexagonally packed single layer.}, and a range of frequencies and accelerations, a fully fluidized, disordered state is observed (see \cite{olafsen98} for representative phase diagrams).   At low vibration amplitudes, the fluctuations in the fluidized state include long-lived, high density, low granular temperature (average kinetic energy of horizontal motion per particle) clusters\cite{olafsen99}.  These fluctuations occur at peak accelerations at or below 1g,  where the dependence of the energy flux from the vibrating plate on the dynamics of the granular layer\cite{geminard03} probably reinforces the tendency of clusters to form due to dissipative interparticle collisions.  

 At still lower accelerations, the fluid becomes unstable to the formation of a `collapse', a hexagonal close packed layer of particles at rest on the plate, which coexists with a surrounding granular fluid.  As with the clustering, the mechanism for the instability can be understood by considering the  energy injection from the plate.  In particular, at low vibration amplitudes it is possible for a particle to remain at rest on the plate, but it is also possible for a particle to bounce periodically with sufficient energy input to overcome dissipative collisions\cite{geminard03,losert99a, urbach01}, even when $\Gamma < 1$.  More generally it has been shown that energy injection that depends on the local kinetic energy can lead to instabilities in the fluidized state\cite{cafiero00}. The collapse-liquid coexistence, and the dependence of the pressure in the liquid state on density, vibration amplitude, and plate material has been studied in detail by G\'eminard and Laroche\cite{geminard04}.  The transition to the fluidized state upon increasing acceleration has been studied in Ref. \cite{losert99a}.

Above about 80\% coverage a fluctuating but hexagonally ordered state is observed for a range of frequencies and vibration amplitudes.  First observed by Pieranski and co-workers\cite{pieranski78,pieranski84},  this phase appears to be separated from the disordered liquid by a continuous phase transition that fits the equilibrium 2D melting scenario surprisingly well\cite{olafsen04}.   In the absence of a confining lid, this phase is only observed at relatively high frequencies.  This may be due to the fact that, for a given acceleration, at higher frequencies the maximum height of the trajectories is lower and therefore the behavior of the granular layer is closer to a true two-dimensional system.  Some evidence for this interpretation comes from the observation that  the phase diagrams at low accelerations measured for two different ball sizes collapses when the frequency axis is scaled by the characteristic frequency $\nu _c =(g/\sigma)^{1/2}$, where $\sigma$ is the ball diameter \cite{urbach01}.  The characteristic time scale, $\nu _c ^{-1}$, is related to the time it takes a ball to fall 1 $\sigma$ due to gravity.  In the next section we show that by confining the layer with a lid while shaking with a high vibration amplitude, many of these complicating gravity-dependent factors can be eliminated, and the resulting behavior bears a remarkable resemblance to that observed in equilibrium systems.

\subsection{The effect of confinement} \label{confinement}
\begin{figure}[h]
\centerline{ }
\center
\includegraphics[width=5cm]{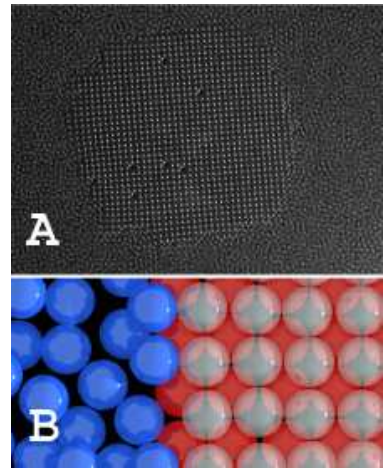}
\caption{{\small{\bf A.} Experimental observation of coexistence of liquid and
    solid phases in a confined granular layer  with $\rho _{2D}=N/N_{max}=0.9$, H = 1.75 $\sigma$, $\nu = 80$Hz, A=0.085$\sigma$.  
The image is an
    average over 1 second. {\bf B.} A three-dimensional rendering of
    particle positions near a solid-liquid interface appearing in 
the computer simulation, for similar parameters.  Fluid-like particles are colored blue, the top layer solid-like particles are transparent and the bottom layer solid-like particles are colored red.}}
\label{coexistence}
\end{figure}

We have previously reported the discovery of a dramatic solid-liquid phase transition that
occurs in the granular layer when the layer is confined with a lid and shaken hard\cite{prevost04}.   Figure \ref{coexistence}A shows an image of the two-phase
coexistence observed in the experiment.  This coexistence represents a steady
state and persists as long as the vibration is maintained. The same
behavior is observed in the simulations, and Fig. \ref{coexistence}B,
a rendering of the simulation data, shows more clearly the structure
of the ordered phase.  It consists of two layers of a square packing,
rather than the hexagonal packing normally expected for hard spheres.

\begin{figure}[h]
\centerline{ }
\center
\includegraphics[width=7cm]{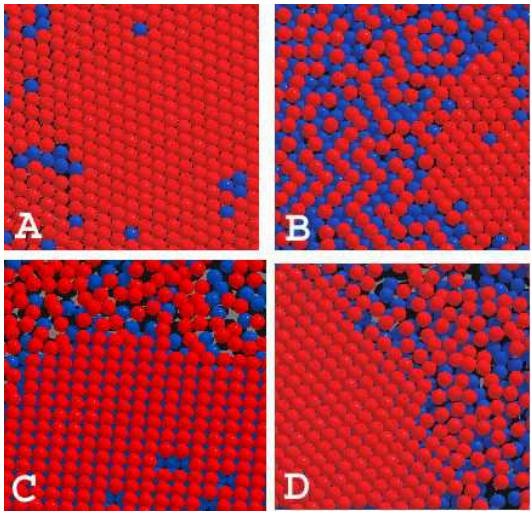}
\caption{{\small Four of the phases encountered as a function of reduced gap height, $h$
in the simulations.  For clarity, particles in the top half of the cell are colored red and particles in the bottom half of the cell are colored blue.  {\bf A.}  $h=0.3\sigma$: Single layer hexagonal.  
{\bf B.} $h=0.5\sigma$: Buckling and single-layer hexagonal coexistence.
{\bf C.} $h=0.7\sigma$: 2-Layer square and liquid coexsitence 
{\bf D.} $h=0.88\sigma$: 2-Layer hexagonal and liquid coexistence.  All simulations are for $\rho_{h}=0.9, A=0.15 \sigma$, and $\nu=75$ Hz.}}
\label{fig:phases}
\end{figure}

This surprising behavior can be understood, at least in part, by
considering the very similar behavior observed in equilibrium hard
sphere colloidal suspensions 
confined in a similarly narrow gap\cite{pansu83}.  
These systems show a complex phase diagram that depends on the
volume fraction of the colloidal particles and the ratio of the
confining gap to the particle 
diameter\cite{pansu83, pieranski83, pansu84,schmidt96, schmidt97,zangi00}. 
The phase diagram can be explained in terms of entropy maximization.
The configuration that gives the particles the most room to rattle
around, within the constraints imposed by the gap, will minimize the
free energy of the system. For a gap of 1.75 $\sigma$,
it turns out that two layer square packing is more efficient than hexagonal and hence has a lower free energy.  As the gap height and volume fraction are varied, the arrangement of the balls which gives the most efficient packing changes.  Figure \ref{fig:phases} shows several of the phase found in our computer simulations of the granular layer with different heights and a fixed volume fraction.  The ordering observed varies from fluid, single layer hexagonal ($\bigtriangleup$), buckling ($b$), two layer squares ($2\Box$), and two layer hexagonal ($2\bigtriangleup$).  There are coexistence regions between neighboring phases where, for example, two hexagonal layers coexist with a fluid ($2\bigtriangleup, f$).

\begin{figure}[h]
\centerline{ }
\center
\includegraphics[width=6.5cm]{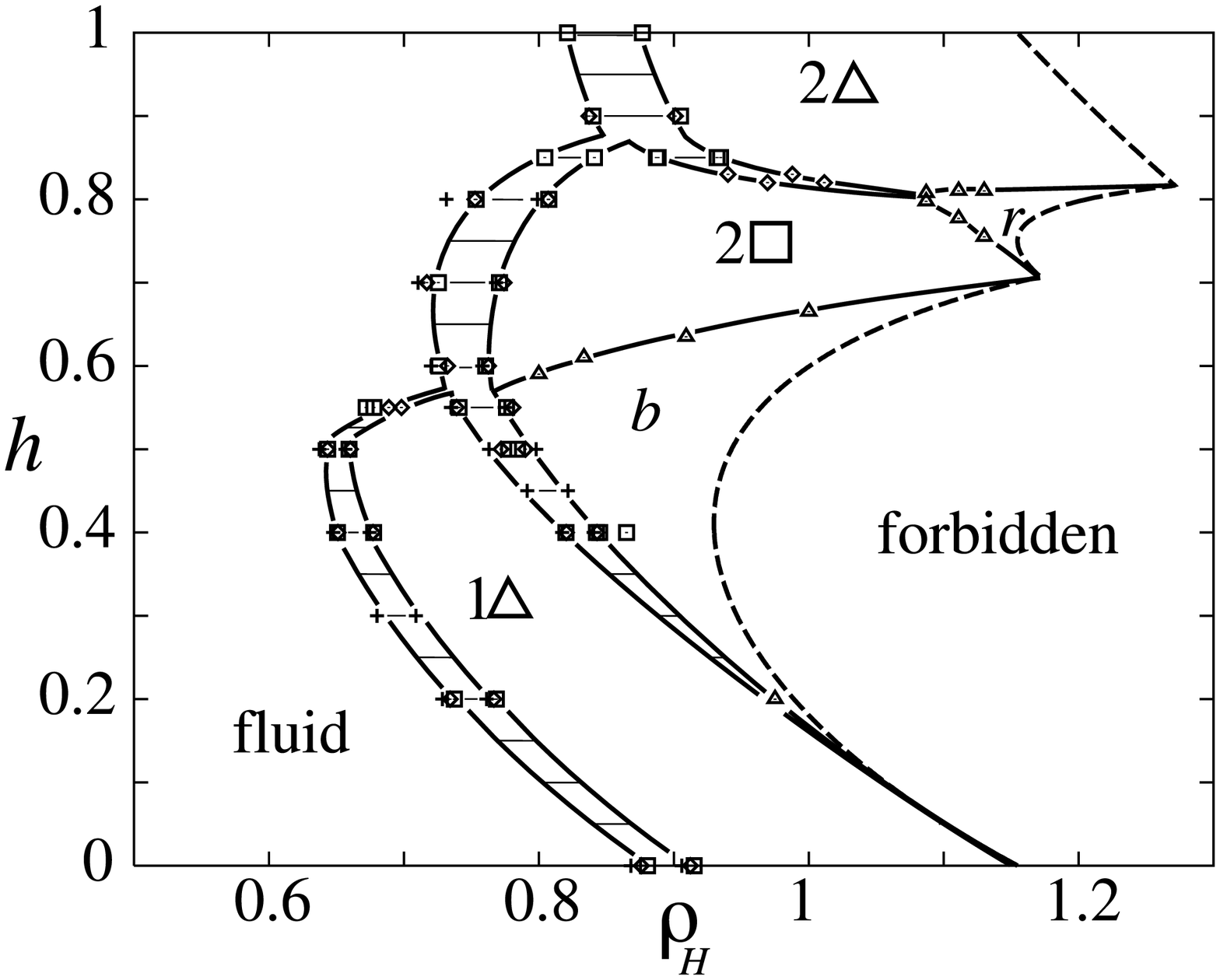}
\includegraphics[width=7cm]{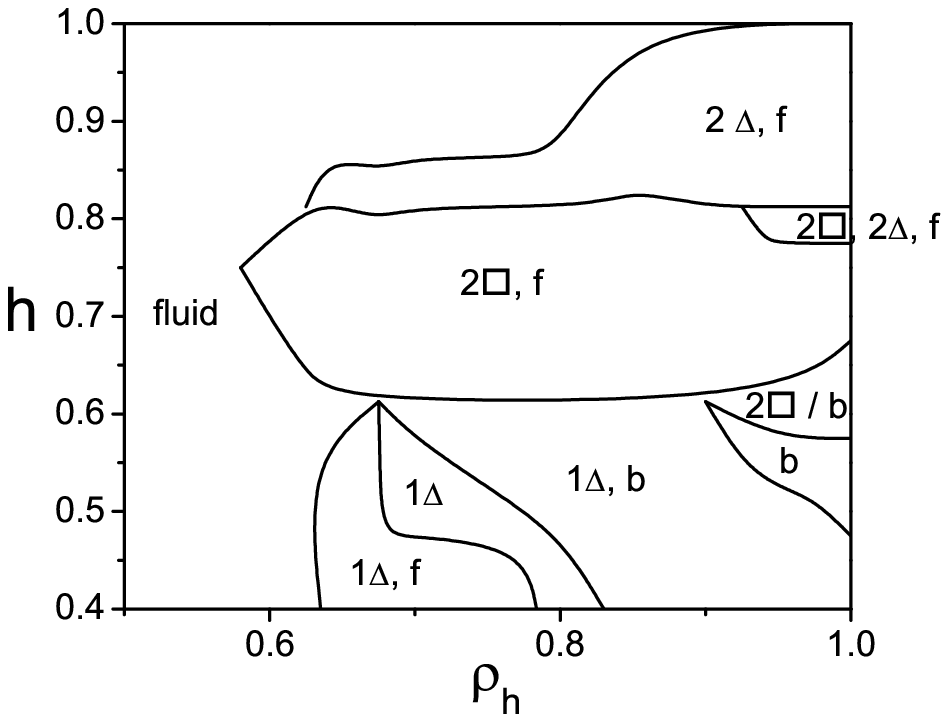}
\caption{{{\bf A.} Phase diagram for the hard sphere colloids, reproduced from Ref. \cite{schmidt97}.  The areas filled in with dashed lines represent coexistence regions of neighboring phases.  {\bf B.} Phase diagram for the granular system.  Here, $A=0.15\sigma$, $\nu=75$Hz, $N=3000$ particles. }}
\label{fig:phasediagram}
\end{figure}

\begin{figure}[h!]
\centerline{}
\center
\includegraphics[width=9cm]{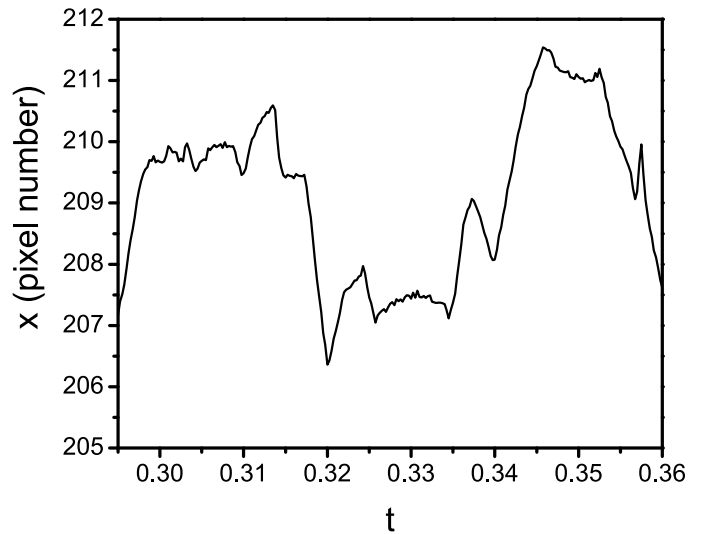}
\caption{\small{Trajectory of single particle in the top layer of the two square layer solid phase.  The experiment was performed using the Cooke 1200HS camera at 4000 frames/second.}}
\label{fig:traj}
\end{figure}

We have mapped out the phase behavior of the granular system by varying the reduced gap height, $h$ (where $h=(H-1)$ and $H$ is the gap height in units of ball diameter, $\sigma$) and reduced volume density, $\rho_{H}$ ($\rho_H=N\sigma^{3}/(SH)$ where $N$ is the number of particles and $S$ is the area of the plate).  The phase diagram for the granular system, derived from the simulations is shown in Fig. \ref{fig:phasediagram}B.  To determine the phase diagram, simulations were performed on a grid of $\rho_h$ and $h$ values which were initially spaced at regular intervals of 0.05 in both variables.  After determining the phase at each ($\rho_h, h$) point, further simulations in the transition regions were performed with a finer grid spacing of 0.02 in both $\rho_h$ and $h$.  Finally, boundary points were interpolated from the simulated data and connected with lines to mark the boundaries.  At values of $\rho_h > 1$ we are able to observe both the $2\Box$ and $2\bigtriangleup$ phases without the presence of the coexisting fluid.  However, for these large values of $\rho_h$ it is difficult to accurately map out a phase diagram as there is little room for the spheres to rearrange after being placed inside the system, so our phase diagram is limited to values of $\rho \le 1$.  For comparison, Fig. \ref{fig:phasediagram}A shows the phase diagram for the colloidal system studied in Ref. \cite{schmidt97}, while the granular phase diagram that we found is presented in Fig. \ref{fig:phasediagram}B.  Overall, the phase diagrams are very similar, with a few notable differences.  The most dramatic difference is the width of the coexistence regions, ($2\Box,$ f) and ($2\bigtriangleup,$ f).  This difference can be understood in terms of the dissipation present in the granular system.  The crystal structure that forms, whether of square or hexagonal symmetry, is of higher density than the surrounding fluid and hence has a higher rate of dissipation.  This higher dissipation causes the crystal particles to cool and the resulting pressure imbalance compresses the crystal.  Because of the compression the crystal that forms in the granular system has a higher density than in the colloidal system, and it therefore requires a much larger increase in overall density to fill the system with the crystal phase.  We have directly measured a significant difference in granular temperature between the coexisting phases in both experiment and computer simulations\cite{prevost04}.   Newer experiments have shown that the dynamics of the spheres inside the crystal is rather complicated.  Fig \ref{fig:traj} shows a typical trajectory for one particle.  The numerous regions where the trajectory is not varying linearly with time, imply that there are frequent non-collisional forces on the particle, such as frictional drag forces with the lid or enduring contacts with neighboring particles.

\begin{figure}[h!]
\centerline{}
\center
\includegraphics[width=9cm]{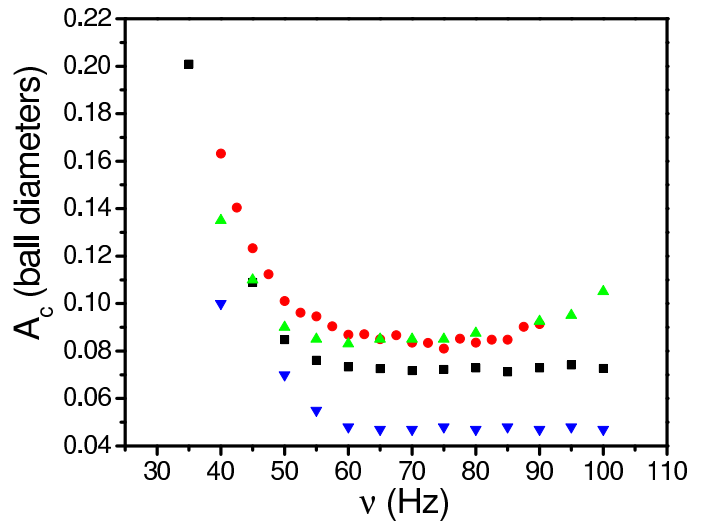}
\caption{\small{ The frequency dependence of the critical amplitude
    necessary to nucleate the solid phase. For all cases, $\rho_{2D}= 0.85$, $H=1.75 \sigma$.  Black squares: steel.  Red
    circles: Brass.  Green and Blue triangles: MD Simulations}}
\label{fig:frequency}
\end{figure}

The similarity between the ordering behavior observed in the granular
and colloidal layers strongly suggests that very similar 
physics underlies both systems.  Unfortunately,
quantitative descriptions of the equilibrium system start with an
analysis of the free energy (which, for hard spheres, depends only on
the entropy).  There is no recipe for adapting this methodology to
strongly non-equilibrium granular dynamics, and a useful analysis of this 
phase behavior will have to take into
account the role of the external driving and energy dissipation.  Figure \ref{fig:frequency}
shows the frequency dependence, in both the experiment and the simulations, of the vibration amplitude necessary to
nucleate a two square layer crystal. (Note that we have plotted the critical value of $A$, the amplitude of the oscillation of the plate position, rather than $\Gamma$.) When the
vibration amplitude is first increased above the critical amplitude, one or more disconnected  crystals 
quickly nucleate from the homogeneous liquid, far from the boundaries of the plate.  The crystals grow and eventually merge until a steady two-phase coexistence
is reached.\footnote{This transition appears to behave like a
hysteretic first-order phase transition, so that a precise critical
amplitude is difficult to define for finite observation times.  We
have found that for moderately slow rates of increase or decrease
the values obtained are only
very weakly dependent on the rate of change of the amplitude.}  
The low frequency
increase in the critical amplitude occurs because the plate
accelerations are not large compared to the Earth's gravity, and the
particles have barely enough energy to reach the second layer.  At high frequencies, the frequency dependence is very weak for steel particles. We have speculated that the formation of the solid phase upon increasing vibration amplitude is a consequence of layer compression\cite{prevost04}.  For brass particles, with a lower coefficient of restitution $e$, the critical amplitude is greater and there is a slow rise in the critical amplitude at higher frequencies.  In the simulations, we observe the same frequency dependence of the critical amplitude and similar dependence on the inelasticity.   These results show that the nonequilibrium effects due to the forcing and the dissipation are an important component of the phase diagram, despite the obvious similarity to that observed in the equilibrium colloidal system.

The phase behavior of highly confined vibrated granular particles shows many parallels with the phase behavior of similarly confined colloidal particles in equilibrium.  This close correspondence suggests that similar driving mechanisms are at work, and that a predictive theory of the phase diagram for this system may be derivable from the equilibrium framework, with suitable modifications to account for the observed dependence on driving and dissipation.

\section{Statistical characterization of the granular gas}
\label{kinetic}
\subsection{Introduction}
\label{kinetic-intro}

While developing a kinetic theory of granular fluids analogous to the kinetic theory of molecular fluids at equilibrium \cite{Enskog, EnskogEn, Chapman} has proven to be very difficult, there has been considerable progress. \cite{Campbell,GoldhirschRev}  For example, the hydrodynamic transport coefficents for granular gases have been calculated for some situations \cite{Goldhirsch, Brey, Garzo}.  The approach taken to develop the hydrodynamic description of granular gases often begins with the solution of the Boltzmann equation in the
spatially homogeneous state, where all gradients are zero. In the
case of undriven granular fluids, this state is the somewhat
artificial `homogeneous cooling state'
which continually loses energy. The Boltzmann
equation for this state was first solved in the work by
Goldshtein and Shapiro \cite{Goldshtein}. Some
recent analytical work \cite{vannoije98} and Monte Carlo simulations
\cite{Montanero} have focused on the non-equilibrium steady state
obtained when the energy supplied by spatially uniform random
forcing is balanced by collisions. In addition, in a more
realistic approach to our experimental system, the boundary
conditions for a vertically vibrated granular fluid have also been
studied \cite{Kumaran,BreyNS,Soto} and the normal fluid state of
the vertically vibrated granular layer has been determined, to
first order in the gradients \cite{BreyNS}. Non-Maxwellian velocity
distributions and velocity correlations have been observed in
these studies and also in a number of experiments and simulations
\cite{olafsen98,olafsen99,losert99,rouyer00,blair01,aranson01,baldassarri01,vanzon04}.
Models incorporating 
fluctuations in the rate of energy input from the driving to
the kinetic energy of the particles, either
through random collisional transfer\cite{barrat01,barrat02} or through
a coupling to the fluctuating
particle velocities\cite{cafiero00}, have succeeded in modelling some
aspects of the observed distributions. 
Nevertheless, the forcing in the experiments is sufficiently
different from that of the theoretical models and a detailed
comparison is difficult. The effect of correlations on the
transport coefficients is still being studied \cite{dufty02,lutsko02,poschel02,alam02, alam02a,dufty01}.
Below we review our results on the statistical characterization of
a vertically driven granular monolayer and present measurements of
velocity autocorrelation functions.

\subsection{Pair Correlation Functions}
\label{position}
Correlations in particle positions are most directly measured by the pair
correlation function, $G(r)$:

\begin{equation}
G(r) = \frac{<\rho(r') \rho(r'+r)>}{<\rho(r')>^2},
\end{equation}

\noindent
where $\rho$ is the particle density and the average is taken over the spatial variable r'. The correlations of a
two-dimensional gas of elastic hard disks in equilibrium 
are due only to geometric
factors of excluded volume and are independent of temperature.  This provides a useful reference point for the measurements in the experimental system.

We have observed three distinct types of behavior of the pair correlation function.  At low vibration amplitudes where strong clustering is observed, $G(r)$ has deeper modulations and a higher value at contact than a system of equilibrium hard disks at the same density\cite{olafsen98}.  The relatively long-lived high density, low granular temperature fluctuations described in Sec. \ref{phases} are presumably responsible for the strong spatial correlations. 

At plate accelerations considerably above 1g in the system without a lid, $G(r)$ loses all
of its structure as the layer goes fully three-dimensional.   The 
change to 3D behavior can be seen
by the increase in the value of G$(r)$ for $r<1$\cite{olafsen99}. (The measured correlation function
includes only horizontal particle separations, so that colliding particles at different heights appear to interpenetrate.)  This
change can affect the dynamics in several ways.  The effective
density is decreased so that excluded volume effects are less
important, the inter-particle collisions can occur at angles closer to
vertical, affecting transfer of energy and momentum from the vertical
direction to the horizontal, and the change in dimensionality itself can have important consequences.

In order to separate these effects
from the direct consequences of increasing the kinetic energy of the gas,
the Plexiglas lid was added to the system, with a gap small enough so that
the particles cannot pass over top of one another, although
enough room remains for collisions between particles at sufficiently
different
heights to transfer momentum from the vertical to the horizontal direction.

 $G(r)$ measured
with the lid on shows that the particle-particle
correlations persist and become independent of $\Gamma$ when the
system is constrained in the vertical direction\cite{olafsen99}. The structure
observed in the correlation function is indistinguishable  from that of an equilibrium elastic hard disk gas at the same two-dimensional density,
indicating that the correlations that exist are due to excluded volume
effects.  From $\Gamma = 1.50$ to $\Gamma = 3.0$, the granular
temperature changes by more than a factor of 2\cite{olafsen99}, yet
there is no detectable change in the pair correlation function.  

\subsection{Velocity Distributions}

The distribution of horizontal velocities of the granular particles is
typically non-Maxwellian. As with the pair correlation function, we have observed a number of different regimes.  At low vibration amplitudes where the clustering is very strong, the velocity distribution has a nearly exponential shape, with a cusp at zero velocity and high velocity tails that are nearly straight on a log-linear plot\cite{olafsen98}.  The broad distribution presumably arises from the fact that the local granular temperature, like the local density, is inhomogeneous in the clustering regime\cite{olafsen99}, and it has been shown very recently that the cusp at zero velocity is likely due to friction with the bottom plate\cite{vanzon04a}.
As the acceleration is increased, the velocity distribution crosses over smoothly to a Maxwellian distribution\cite{olafsen99}.  Identical behavior was observed in molecular dynamics simulations designed to mimic the experimental situation\cite{nie00}.  In addition, those simulations showed that, at low accelerations,  the average kinetic energy (granular temperature) in the vertical motion was several times larger than that in the horizontal.  The approach to Maxwellian horizontal velocity distributions coincides with a dramatic reduction in the anisotropy.  This is consistent with the observation that non-Maxwellian features in the velocity distribution can arise from energy transfer through collisions in strongly anisotropic situations\cite{barrat02}.  The expansion of the layer in the vertical direction as the acceleration is increased presumably also results in more interparticle collisions relative to particle-plate collisions, and this ratio may play an important role in determining the form of the velocity distributions\cite{vanzon04}.

When the expansion of the layer is eliminated by adding the lid, the crossover to Maxwellian distributions is not observed\cite{olafsen99}. As with $G(r)$, the velocity distribution becomes essentially independent of the acceleration for values above roughly 1.5 g.  However, whereas $G(r)$ becomes essentially indistinguishable from that observed in equilibrium systems, the velocity distribution has a cusp at zero velocity and high-velocity tails that are significantly broader than Gaussian tails.  

When the smooth bottom plate is replaced by the rough plate, the cusp at zero velocity disappears, but the broad high-velocity tails remain, at least for moderate densities\cite{prevost02}.  At very low densities, however, the distributions approach a Maxwellian distribution.   Measurements at low densities (typically below $\rho_{2D}=0.25$) are not possible on the flat plate, because the interparticle collisions are too infrequent and all horizontal motion ceases.  By contrast, every collision with the rough plate injects energy directly into the horizontal, as well as vertical, ball motion.  As a result even a single ball on the plate will exhibit apparently random changes in direction, much like the random forcing of some models of heated granular fluids\cite{vannoije98,vannoije99,puglisi99,pagonabarraga01}.  As described below,  the behavior of the velocity autocorrelation function suggests that the scattering of the ball with the rough plate acts as a viscous drag.  Thus the Maxwellian velocity distributions at low densities may result from Langevin-like dynamics, with the plate providing both the random forcing and the drag.  As the density is increased, both the drag and the interparticle collisions will affect the velocity distribution, a situation modeled by Puglisi, et al.\cite{puglisi99}.  However the granular temperature is always anisotropic in the quasi-2D layers, so the energy transfer from the vertical to the horizontal components of the velocity will affect the velocity distribution on the rough plate as well\cite{barrat02}.  A somewhat similar geometry was investigated in Ref. \cite{baxter03}, where a full layer of heavy two-particle chains played the role of the rough plate.  This provides a dynamic bottom layer, in contrast to our rough plate, where the bottom layer of balls is glued to the oscillating flat surface. The velocity distributions appear to be closer to a Maxwellian than those we have reported (compare Fig. 5 of Ref. \cite{prevost02} with Fig. 1 of Ref. \cite{baxter03}), but there are a number of differences between the experiments, including the relative sizes of the balls in the two layers and the material properties of the particles, so that it is not clear if the quantitative differences are due to the motion of the bottom layer or to other effects.  

\subsection{Velocity autocorrelation function}

\begin{figure*}[h]
\begin{center}
\includegraphics[width=18cm]{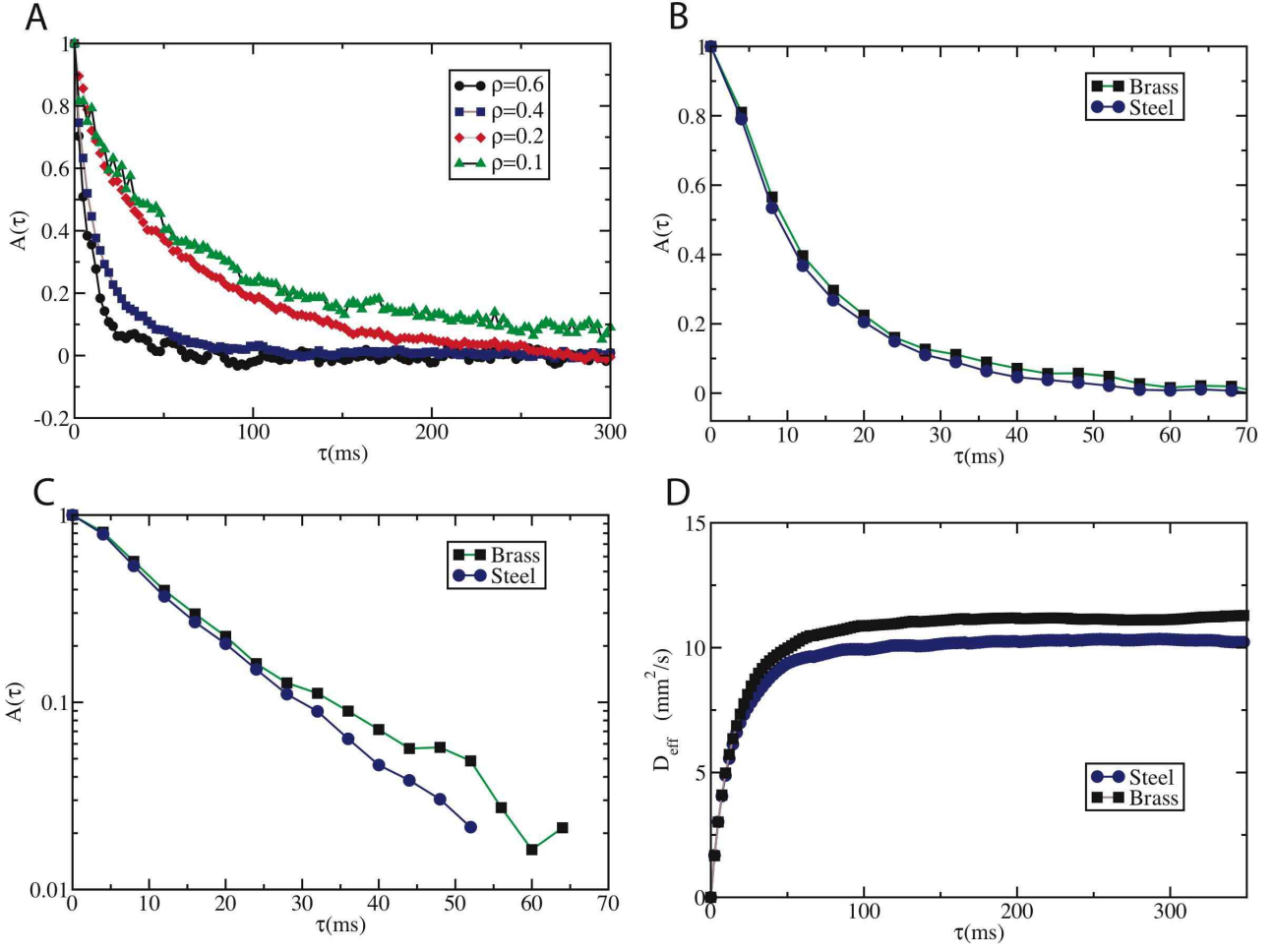}
\caption{{\bf Autocorrelation and self-diffusion on the rough plate}  {\bf A.} $A(\tau)$ at different values of $\rho_{2D}$  at $\Gamma = 1.5$ {\bf B.}  $A(\tau)$ for steel and brass at $\rho_{2D}=0.6$ and $\Gamma = 1.39$ (steel) and $\Gamma = 1.57$ (brass).  The vibration amplitude was chosen so that the granular temperature was the same in both cases. {\bf C.} Same data as in {\bf B} on a log-linear  plot. {\bf D.} Mean square displacement divided by time, for the same conditions as {\bf B}.   All data were collected for  $\nu$=60 Hz.}
\label{autocorr}
\end{center}
\end{figure*}

As described above, continuum models of granular
flows need to account for the presence of correlations, which  can be seen directly in the
self-diffusion of individual particles.  We have measured the normalized velocity
autocorrelation function,   
\begin{equation}
A(\tau) = \frac{<v_i(t) v_i(t+\tau)>}{<v_i(t)>^2},
\label{eq:two}
\end{equation}
where $v_i$ is one component of the horizontal velocity and the brackets represent the time average.  While in principle the hexagonal symmetry of the rough plate could produce an asymmetry in the autocorrelation function, none is observed, so the data shown is the average of the result for each component separately.
Fig. \ref{autocorr}a shows  $A(\tau)$ measured on the rough plate at different densities at the same plate acceleration.  As expected, the autocorrelation function decays more slowly as the density decreases because of the longer time between interparticle collisions.  At the lowest densities, however, the decay rate becomes almost independent of the density, suggesting that scattering from the rough plate is the dominant mechanism of momentum decay at those densities.

The autocorrelation functions for brass and steel spheres at a density of $\rho _{2D}= 0.6$ are compared in  Fig. \ref{autocorr}b. We used a slightly higher plate acceleration for steel than for the more dissipative brass ($\Gamma=1.57$ vs.    $\Gamma=1.39$) to produce the same horizontal granular temperature.  The autocorrelation functions show an approximately exponential decay at short times, as it can be seen more clearly in Fig. \ref{autocorr}c, when plotted on a log-linear plot.  The decay rate is about a factor of two faster than that calculated from kinetic theory for a liquid at the same density and temperature, presumably due to the added effect of the scattering with the rough plate discussed above.  At longer times there is clearly upward curvature on the log-linear plot. 
This slow temporal decay is likely
related to the equal time spatial correlations reported
previously\cite{prevost02} and described below. (Although long time tails in the velocity
autocorrelation function are present in dense elastic fluids as
well\cite{hansen86}, those effects are smaller than what we observe in the granular fluid ). The slower decay is more dramatic for the brass, probably because the  
spatial velocity correlations, which are a consequence of the forcing and dissipation, are stronger, but we have not confirmed this.

The diffusion coefficient is directly related to the integral of the
velocity autocorrelation function, but it can be measured most
readily from a plot such as Fig. \ref{autocorr}d, which shows the mean
squared particle displacement divided by time for the brass and
steel spheres. For normal diffusive behavior, this curve will reach a
constant value (proportional to the diffusion coefficient) at long times.
Both brass and steel clearly exhibit normal diffusion, but the diffusion coefficient for brass is higher than that of steel.  In the absence of correlations, the diffusion
coefficient 
given by hard-disk kinetic theory depends only on the mean free path, which
depends only on the density, and the granular temperature. Thus the enhanced diffusion in brass is a direct demonstration of the effect of forcing and dissipation on transport properties in granular gasses.

\subsection{Spatial velocity correlations}
The assumption of `molecular chaos', that the velocities of colliding particles are uncorrelated, 
is a crucial approximation normally used to solve the
Boltzmann equation and to calculate other fundamental quantities
in the kinetic theory of liquids.  The presence of spatial velocity correlations can significantly affect collisional transport of mass, energy, and momentum. \cite{Garzo, Ernst}

The longitudinal and transverse velocity correlations, $C_{||}$
and $C_{\bot}$, respectively,  are calculated by
\begin{eqnarray}\label{velocitycorrdef}
C_{||,\bot}(r)= \sum_{i\neq j}^{N_r} v_i^{||,\bot}
v_j^{||,\bot}/N_r, \nonumber
\end{eqnarray}
where the sum runs over the $N_r$ pairs of particles separated by
a distance $r$, $v_i^{||}$ is the projection of ${\bf
v}_i$ along the line connecting the centers of particles $i$ and
$j$, and $v_i^{\bot}$ is the projection perpendicular to that
line.  (Only the horizontal components of the velocity are measured).  Measurements on the smooth plate revealed surprising strong {\em negative} values for $C_{||}(r)$, indicating that the particles in the granular gas exhibit a tendency to move apart from one another\cite{prevost02}.    On the smooth plate, there is always more energy in the vertical components of the velocity than in the horizontal components, because energy is injected exclusively into the vertical motion.  This energy gets transferred to the horizontal components during collisions, so that the particles have a tendency to `fly apart' from collisions, resulting in the observed anti-correlated velocities.   

Measurements on the rough plate, by contrast, show $C_{||,\bot}(r)$ everywhere positive, with an exponential decay that is approximately independent of density.  A model system with random forcing also shows positive correlations, but with a slower decay that depends on the density of the granular gas\cite{vannoije98,vannoije99,pagonabarraga01,moon01}.    This difference may be due to the scattering of the balls by the rough plate.  These collisions do not conserve momentum, and are therefore probably more effective than ball-ball collisions at dissipating the velocity correlations which arise from the effects of the random forcing.  The dramatic change in the velocity correlations when the bottom surface is changed provides a direct demonstration that the kinetics of the granular gas are very sensitive to the forcing mechanism.

\section{Conclusion}
\label{conclusion}

The surprisingly rich behavior of the simple system described in this paper is in many ways emblematic of the opportunities and challenges presented by granular flows more generally.  There are situations where the statistical mechanics of thin vibrated layers is essentially similar to equilibrium systems, such as the existence of nearly Maxwellian velocity distributions at high vibration amplitudes in the absence of confinement, or at low densities on the rough plate.  On the other hand, there are situations where the driving and the dissipation produce strongly non-equilibrium behavior, such as strongly non-Maxwellian velocity distributions and significant velocity correlations, evidence for violations of molecular chaos.  The observation of a complex phase diagram that is clearly similar to a well understood equilibrium system suggests that entropy maximization, the mechanism that drives the  transitions in equilibrium, can operate far from equilibrium.  Adapting that formalism to account for the observed effects of forcing and dissipation remains an open challenge.

\subsection{Acknowledgments}  The authors are thankful to Leon Der for contributions to the construction of experimental apparatus and to Jeff Olafsen and Heather Deese for their contributions to some of the results reviewed in this paper.
This work was supported by grant
DMR-9875529 and DMR-0094178 from the NSF and by grant NNC04GA63G from NASA. D.A.E. acknowledges support from the Alfred P. Sloan foundation.
F. V. R. is supported by the Spanish Government as an FPI-MECD and SEEU Fellow (Secretary of State of Education and Universities, ref. GT2002-0023) and also through contract BFM-2003-01739. 

\end{document}